\documentclass[superscriptaddress,nofootinbib,twocolumn]{revtex4-2}
\usepackage{amssymb}
\usepackage{amsmath}
\usepackage{graphicx}
\usepackage{mathdots}
\usepackage{slashed}
\usepackage{verbatim}
\usepackage{cancel}
\usepackage{float}
\usepackage[colorlinks=true,linkcolor=blue,citecolor=blue,urlcolor=blue]{hyperref}

\makeatletter
\def\@hangfrom@section#1#2#3{\@hangfrom{#1#2}#3}%\MakeTextUppercase{#3}}%
\def\@hangfroms@section#1#2{#1#2}%\MakeTextUppercase{#2}}%
\makeatother

\newcommand{\be}{\begin{equation}}
\newcommand{\ee}{\end{equation}}
\newcommand{\bea}{\begin{eqnarray}}
\newcommand{\eea}{\end{eqnarray}}

\makeatletter
\def\@ssect@ltx#1#2#3#4#5#6[#7]#8{%
  \def\H@svsec{\phantomsection}%
  \@tempskipa #5\relax
  \@ifdim{\@tempskipa>\z@}{%
    \begingroup
      \interlinepenalty \@M
      #6{%
       \@ifundefined{@hangfroms@#1}{\@hang@froms}{\csname @hangfroms@#1\endcsname}%
       {\hskip#3\relax\H@svsec}{#8}%
      }%
      \@@par
    \endgroup
    \@ifundefined{#1smark}{\@gobble}{\csname #1smark\endcsname}{#7}%
  }{%
    \def\@svsechd{%
      #6{%
       \@ifundefined{@runin@tos@#1}{\@runin@tos}{\csname @runin@tos@#1\endcsname}%
       {\hskip#3\relax\H@svsec}{#8}%
      }%
      \@ifundefined{#1smark}{\@gobble}{\csname #1smark\endcsname}{#7}%
      \addcontentsline{toc}{#1}{\protect\numberline{}#8}%
    }%
  }%
  \@xsect{#5}%
}%
\makeatother

\begin{document}

%===================================================
\title{Iterative Gauging is Deconstruction}
%===================================================

\author{Sebasti\'an Franco}\email{sfranco@ccny.cuny.edu}
\affiliation{Physics Department, The City College of the CUNY\\
	160 Convent Avenue, New York, NY 10031, USA}
\affiliation{Physics Program and Initiative for the Theoretical Sciences\\
	The Graduate School and University Center, The City University of New York\\
	365 Fifth Avenue, New York NY 10016, USA}
\author{Diego Rodr\'iguez-G\'omez}\email{d.rodriguez.gomez@uniovi.es}
\affiliation{Department of Physics, Universidad de Oviedo \\  
C/ Federico Garc\'ia Lorca  18, 33007  Oviedo, Spain}
\affiliation{Instituto Universitario de Ciencias y Tecnolog\'ias Espaciales de Asturias (ICTEA) \\
 C/~de la Independencia 13, 33004 Oviedo, Spain.}

\begin{abstract}
A recent construction showed that iteratively gauging symmetries of a quantum field theory can generate an emergent extra spatial dimension. Here we identify the precise mechanism underlying this phenomenon: it is dimensional deconstruction, the procedure by which an extra dimension is encoded in the structure of a quiver gauge theory. We demonstrate this by showing that the standard deconstruction action, upon dualizing the Goldstone fields on the Higgs branch, is exactly equivalent to the action produced by iterative gauging. The dictionary is transparent: quiver nodes correspond to gauge fields, and quiver links correspond to the gauge fields introduced to couple to magnetic symmetries. We further show that starting by gauging the electric rather than the magnetic symmetry produces the deconstructed theory written in the dual variables, with all coupling constants inverted.

\end{abstract}

\maketitle

\makeatletter
\let\toc@pre\relax
\let\toc@post\relax
\makeatother 

%===================================================
\section{Introduction}
%===================================================

The idea that an extra spatial dimension can emerge from purely lower-dimensional physics has a long and fruitful history. A concrete realization is provided by \emph{dimensional deconstruction} \cite{Arkani-Hamed:2001kyx} (see also \cite{Arkani-Hamed:2001nha,Arkani-Hamed:2001wsh}): a quiver gauge theory on the Higgs branch develops a mass spectrum for its gauge fields that is indistinguishable from that of a theory with a latticized extra dimension. The key signature is a nearest-neighbour mass matrix of the harmonic-oscillator type, which reproduces Kaluza--Klein towers in the continuum limit.

More recently, a different route to emergent dimensions was proposed in Ref. \cite{Rubio:2024rau} (see also \cite{Blanik:2025uvv,GarreRubio:2025dipoles,Blanik:2025nonabelian}): starting from a simple abelian gauge theory and repeatedly gauging its global symmetries, an extra dimension appears to grow iteration by iteration. This was demonstrated by explicit calculation in several systems, but the underlying reason was left implicit.

In this work we explain that reason. We show that the iterative gauging of Ref. \cite{Rubio:2024rau} \emph{is} deconstruction, connected by a standard duality transformation. The two descriptions are not merely analogous: they are related by dualizing the Goldstone scalar fields that appear on the Higgs branch of the quiver. This identification is exact and holds for arbitrary, inhomogeneous coupling constants, which correspond to non-uniform latticizations of the extra dimension. We also show that choosing to gauge the electric rather than the magnetic symmetry at each step produces an equivalent construction in a dual frame, with couplings inverted.

%===================================================
\section{Deconstruction}
%===================================================

\begin{figure}[H]
    \centering
    \includegraphics[width=6cm]{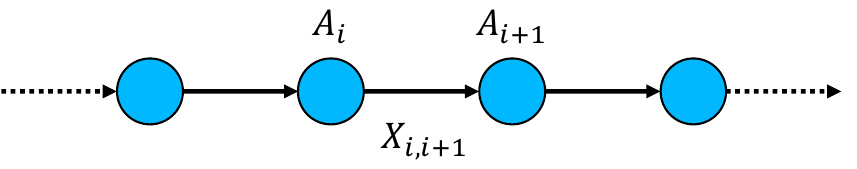}
    \caption{Deconstruction using a linear quiver.}
    \label{quiver_deconstruction}
\end{figure}

We consider a chain of $U(1)$ gauge groups labelled by $i$, connected with complex scalars $X_{i,i+1}$ with charge -1 under a $U(1)$ gauge group $i$ and charge +1 under a $U(1)$ gauge group  $i+1$. Such a theory can be represented by a quiver diagram like the one shown in Figure \ref{quiver_deconstruction}.
The action is

\begin{equation}
S = \int \sum_i\, |dX_{i,i+1}-i(A_{i+1}-A_i)X_{i,i+1}|^2+\frac{1}{2\,g_i^2}|dA_i|^2\,.
\end{equation}
The index $i$ is an index in theory space which gets transmuted into a spacetime index. To see this, we move to the Higgs branch where we set  $X_{i,i+1}=v_{i,i+1}\,e^{i\theta_{i,i+1}}$, being $v_{i,i+1}$ the (frozen) VEVs and $\theta_{i,i+1}$ the (fluctuating) field. Then, the effective action to use is

\begin{equation}
\label{S}
S = \int \sum_i\, v_{i,i+1}^2\,|d \theta_{i,i+1}-A_{i+1}+A_i|^2+\frac{1}{2\,g_i^2}|dA_i|^2\,.
\end{equation}

For the simplest choice $v_{i,i+1}=v$ and $g_i=g$, upon gauge-fixing, we can write

\begin{equation}
S = \int \sum_i\, \frac{1}{2\,g^2}|dA_i|^2+\frac{1}{2} A_i M_{ij} A_j\,,
\end{equation}
with

\begin{equation}
M =
2v^2 \begin{pmatrix}
2 & -1 & 0 & \cdots & -1 \\
-1 & 2 & -1 & \ddots & \vdots \\
0 & -1 & 2 & \ddots & 0 \\
\vdots & \ddots & \ddots & \ddots & -1 \\
-1 & \cdots & 0 & -1 & 2
\end{pmatrix}\,.
\label{matrix_M}
\end{equation}
We see the mass-matrix of the chain of harmonic oscillators, which is the hallmark of a homogeneously latticized extra dimension growing through dimensional deconstruction \cite{Arkani-Hamed:2001kyx}.\footnote{The explicit form of \eqref{matrix_M}, in particular its off-diagonal corner entries, corresponds to a circular quiver. However, whether the quiver is closed into a circle or not plays no role in our discussion.} From this point of view, generic $v_{i,i+1}$ and $g_i$ correspond to generic latticizations of the extra dimension.

%===================================================
\subsection{Dualizing the Goldstone fields}
%===================================================

Let us come back to \eqref{S} and instead dualize $\theta_{i,i+1}$. Introduce $d\theta_{i,i+1}=F_{i,i+1}$ with $F_{i,i+1}$ a 1-form, and impose that $dF_{i,i+1}=0$ through a Lagrange multiplier\footnote{We work in Euclidean signature, so BF couplings carry an explicit factor of $i$. In addition, in the following we will not keep careful track of overall numerical factors such as factors of $2$ or $\pi$, as they are irrelevant for the present discussion.}

\begin{align}
\label{lagrangemultiplier}
S = & \int \sum_i\, v_{i,i+1}^2\,|F_{i,i+1}-A_{i+1}+A_i|^2+\frac{1}{2\,g_i^2}|dA_i|^2\nonumber \\ &+i\,F_{i,i+1}\wedge dH_{i,i+1}\,,
\end{align}
where $H_{i,i+1}$ is a $d-2$-form. Integrating out $F_{i,i+1}$ gives

\begin{align}
\label{SAH}
S =& \int \sum_i\, \frac{1}{2\,v_{i,i+1}^2}|dH_{i,i+1}|^2+\frac{1}{2\,g_i^2}|dA_i|^2+\nonumber \\ & i\,(A_{i+1}-A_i)\wedge dH_{i,i+1}\,.
\end{align}

The dualized action \eqref{SAH} makes the underlying quiver structure particularly transparent: the gauge fields $A_i$ are naturally associated to nodes, while the fields $H_{i,i+1}$ are naturally associated to links. More importantly, this representation provides the bridge between deconstruction and iterative gauging. As we now show, \eqref{SAH} is precisely the action generated by the iterative-gauging procedure of \cite{Rubio:2024rau}.

%===================================================
\section{Deconstruction from iterative gauging}
%===================================================

We now show that the action \eqref{SAH} can be produced by a bottom-up iterative gauging procedure. The algorithm is as follows

\begin{enumerate}
\item \textbf{Seed theory:} start with a Maxwell field

\begin{equation}
S=\int \frac{1}{2\,g_1^2} |dA_1|^2\,.
\end{equation}

This has an electric 1-form symmetry associated to the shift of the gauge field $dA_1\rightarrow dA_1+a$ with $da=0$ whose current is $j^A_e=dA_1$ and a magnetic $d-3$-form symmetry associated to the shift of the dual gauge field $\tilde{A}_1\rightarrow \tilde{A}_{1}+\tilde{a}$ where $d\tilde{A}_{1}=\star dA_1$ and  $d\tilde{a}=0$. The magnetic current is $j^A_m=d\tilde{A}_1=\star dA_1$.

\item \label{2} \textbf{Gauge the magnetic $d-3$-form symmetry:} we gauge the $d-3$-form symmetry through a $d-2$ form gauge field $H_{1,2}$

\begin{equation}
S =\int \frac{1}{2\,g_1^2} |dA_1|^2+\frac{1}{2\,v_{1,2}^2}|dH_{1,2}|^2+i\,H_{1,2}\wedge dA_1\,,
\end{equation}
where we included the kinetic terms for the added gauge field. In addition to the electric 1-form symmetry associated to the shift of the gauge field $dA_1\rightarrow dA_1+a$ with current is $j^A_e=dA_1$, the theory has a 0-form shift symmetry $\tilde{H}_{1,2}\rightarrow \tilde{H}_{1,2}+\tilde{h}$, where $d\tilde{H}_{1,2}=\star dH_{1,2}$ and $d\tilde{h}=0$. The current for this symmetry is $j^H_m=d\tilde{H}_{1,2}=\star dH_{1,2}$.

\item \label{3} \textbf{Gauge the emergent 0-form symmetry:} we gauge the 0-form symmetry through a 1-form gauge field $A_2$. This amounts to an extra coupling $i\,A_2\wedge dH_{1,2}$. Including the kinetic terms for the added gauge fields, after straightforward manipulations, one obtains

\begin{align}
S=& \int \frac{1}{2\,g_1^2} |dA_1|^2+\frac{1}{2\,g_2^2} |dA_2|^2+\frac{1}{2v_{1,2}^2}|dH_{1,2}|^2\nonumber \\ &+i\,H_{1,2}\wedge d(A_1-A_2) \,.
\end{align}

This theory exhibits, associated to $A_2$, a new magnetic symmetry with current $j_m^A=dA_2$.

\item \textbf{Iterate:} we now go back to step \ref{2} and gauge the magnetic symmetry for $A_2$. Doing this recursively we find

\begin{align}
S =& \int \sum_i\, \frac{1}{2v_{i,i+1}^2}|dH_{i,i+1}|^2+\frac{1}{2\,g_i^2}|dA_i|^2\nonumber \\ & +i\,(A_{i+1}-A_i)\wedge dH_{i,i+1}\,.
\end{align}

\end{enumerate}

Thus, from iterated gauging we recover the lagrangian describing the deconstructed extra dimension, with generic $v_{i,i+1}$, $g_i$ corresponds to a generic latticization. This is essentially the procedure introduced in \cite{Rubio:2024rau}, where an extra dimension arises from iterative gaugings. 

It is interesting to look in closer detail to the emergent 0-form symmetry gauged in step \ref{3}. Recall that the symmetry shifts $\tilde{H}_{i,i+1}\rightarrow \tilde{H}_{i,i+1}+\tilde{h}_i$, where $d\tilde{H}_{i,i+1}=\star dH_{i,i+1}$. Hence, combining with \eqref{lagrangemultiplier}, we roughly have $d\tilde{H}_{i,i+1}\sim d\theta_{i,i+1}-A_{i+1}+A_i$. Thus, the shift symmetry $\tilde{H}_{i,i+1}\rightarrow \tilde{H}_{i,i+1}+\tilde{h}_i$ gauged in step \ref{3} is naturally identified with the shift symmetry $\theta_{i,i+1}\rightarrow \theta_{i,i+1}+\tilde{h}_i$, namely, the $U(1)$ rotation of the bifundamental between gauge group $i$ and $i+1$. Hence, our iterative gauging is literally reconstructing the extra dimension by assembling a quiver: sequentially adding the $H$ corresponding to the bifundamentals and gauging their $U(1)$ 0-form symmetry.

%===================================================
\subsection{Electric gauging}
%===================================================

Starting with the Maxwell field, we gauged the magnetic symmetry to ignite our iterative gauging. It is natural to ask what would have happened if we gauged the 1-form electric symmetry. In that case we would have found

\begin{equation}
S=\int \frac{1}{2g_1^2}|dA_1+\tilde{H}_{1,2}|^2+\frac{v_{1,2}^2}{2}|d\tilde{H}_{1,2}|^2\,,
\end{equation}
with $\tilde{H}_{1,2}$ a 2-form gauging the 1-form electric symmetry (the choice of name will become clear momentarily) whose kinetic term with a judicious choice of coupling for latter convenience is included. We now have a magnetic symmetry for $\tilde{H}_{1,2}$ with current $\star d\tilde{H}_{1,2}$. Introducing a $d-3$-form gauge field $C_2$ we gauge this symmetry as

\begin{align}
S=&\int \frac{1}{2g_1^2}|dA_1+\tilde{H}_{1,2}|^2+\frac{v_{1,2}^2}{2}|d\tilde{H}_{1,2}|^2+\frac{g_2^2}{2} |dC_2|^2\nonumber \\ & +i dC_2\wedge \tilde{H}_{1,2}\,.
\end{align}
As usual, we have included the kinetic term for the new gauge field and again chosen judiciously the definition for its coupling for future convenience. Dualizing $C_2$ in the standard way we find

\begin{equation}
S=\int \frac{1}{2g_1^2}|dA_1+\tilde{H}_{1,2}|^2+\frac{1}{2g_2^2} |dA_2+\tilde{H}_{1,2}|^2+\frac{v_{1,2}^2}{2}|d\tilde{H}_{1,2}|^2\,.
\end{equation}
Let us now dualize $A_i$. When the dust settles one finds

\begin{align}
S=&\int \frac{1}{2\tilde{g}_1^2}|d\tilde{A}_1|^2+\frac{1}{2\tilde{g}_2^2} |d\tilde{A}_2|^2+\frac{1}{2\tilde{v}_{1,2}^2}|d\tilde{H}_{1,2}|^2 \nonumber \\  & +i\,\tilde{H}_{1,2}\wedge d(\tilde{A}_1-\tilde{A}_2)  \,,
\end{align}
where $\tilde{g}_i=g_i^{-1}$, $\tilde{v}_{1,2}=v_{1,2}^{-1}$. This is of the form of the action found upon magnetic gauging but in the dual gauge fields. Just as before, in order to find an iterative structure, we should now gauge the magnetic symmetry associated to $\tilde{A}_2$. Adding a gauge field and its kinetic term

\begin{align}
S=&\int \frac{1}{2\tilde{g}_1^2}|d\tilde{A}_1|^2+\frac{1}{2\tilde{g}_2^2} |d\tilde{A}_2|^2\nonumber \\ & +\frac{1}{2\tilde{v}_{1,2}^2}|d\tilde{H}_{1,2}|^2+\frac{1}{2\tilde{v}_{2,3}^2}|d\tilde{H}_{2,3}|^2 \nonumber \\  & +i\,\tilde{H}_{1,2}\wedge d(\tilde{A}_1-\tilde{A}_2)+i\,\tilde{H}_{2,3}\wedge d\tilde{A}_2 \,.
\end{align}
Further gauging of the --now magnetic-- symmetry associated to $\tilde{H}_{2,3}$ with a $d-3$-form $\tilde{A}_3$ produces

\begin{align}
S=&\int \frac{1}{2\tilde{g}_1^2}|d\tilde{A}_1|^2+\frac{1}{2\tilde{g}_2^2} |d\tilde{A}_2|^2+\frac{1}{2\tilde{g}_3^2} |d\tilde{A}_3|^2 \nonumber \\  & +\frac{1}{2\tilde{v}_{1,2}^2}|d\tilde{H}_{1,2}|^2+\frac{1}{2\tilde{v}_{2,3}^2}|d\tilde{H}_{2,3}|^2 \nonumber \\  & +i\,\tilde{H}_{1,2}\wedge d(\tilde{A}_1-\tilde{A}_2)+i\,\tilde{H}_{2,3}\wedge d(\tilde{A}_2-\tilde{A}_3) \,.
\end{align}
 Iterating we now find
 
 \begin{align}
 \label{Selectricgauging}
S=&\int \sum \frac{1}{2\tilde{g}_i^2}|d\tilde{A}_i|^2 +\frac{1}{2\tilde{v}_{i-1,i}^2}|d\tilde{H}_{i-1,i}|^2 \nonumber \\  & +i\,\tilde{H}_{i-1,i}\wedge d(\tilde{A}_{i-1}-\tilde{A}_i) \,.
\end{align}
Thus we see the natural iterative construction is still in terms of magnetic symmetries. Yet, the initial electric gauging imposes a dualization, so that the final outcome is a deconstructed the action written in terms of the gauge fields dual to those appearing in \eqref{SAH}. 

Amusingly, if $d\geq 4$, this action might be re-written as

\begin{equation}
\label{Sdual}
S = \int \sum_i\, \tilde{v}_{i,i+1}^2\,|d \tilde{\theta}_{i,i+1}-\tilde{A}_{i+1}+\tilde{A}_i|^2+\frac{1}{2\,\tilde{g}_i^2}|d\tilde{A}_i|^2\,,
\end{equation}
 where $\tilde{\theta}_{i,i+1}$ are $d-4$-forms. Dualizing these, one easily shows that  \eqref{Sdual} becomes \eqref{Selectricgauging}. Note that in the particular case of $d=4$ the $\tilde{\theta}_{i,i+1}$ are scalars and the $\tilde{A}_i$ gauge fields, so that \eqref{Sdual} is analogous to \eqref{S}.

%===================================================
\section{Discussion}
%===================================================

We have shown that dimensional deconstruction is the universal mechanism behind iterative gauging. The two are related by a standard Goldstone duality, and the dictionary is exact: quiver nodes map to 1-form gauge fields $A_i$, and quiver links map to the $(d{-}2)$-form fields $H_{i,i+1}$ introduced to gauge the magnetic symmetries. The procedure of Ref.~\cite{Rubio:2024rau}, which builds up an extra dimension step by step, is precisely the process of assembling the quiver one bifundamental at a time.

This connection has several implications. First, the generality of the iterative gauging construction inherits all the power of deconstruction: the resulting theory approximates a genuine Kaluza--Klein extra dimension in the limit of many nodes with small lattice spacing. A natural freedom in the iterative gauging construction concerns the boundary conditions imposed in theory space after $N$ steps. For simplicity, we restrict attention to periodic boundary conditions in theory space, identifying $A_{N+1}\equiv A_1$ (and similarly for the corresponding link fields), corresponding to the standard scenario of a deconstructed circle. More general boundary conditions could be imposed, potentially leading to non-trivial boundary sectors, in analogy with the boundary structures that arise in iterative gauging constructions. We leave a detailed investigation of such possibilities for future work. 

Second, the dual construction obtained by starting with gauging the electric rather than the magnetic symmetry shows that iterative gauging is not tied to a single duality frame — it works equally well in the dual description, producing theories with inverted couplings. In this respect, it should be stressed that the initial choice of gauging the electric symmetry forces the transformation into the dual frame. Yet, the recursive structure of the iterative gauging generating the extra dimensions is in both cases tied to magnetic gaugings. The electric-gauging construction also suggests a new perspective on the emergent dimension. Magnetic gauging builds the deconstructed theory directly in terms of the original 1-form gauge fields, whereas electric gauging inevitably drives the construction into the dual frame, yielding an emergent dimension described by $(d-3)$-form gauge fields. The existence of these two iterative constructions hints that the same emergent geometry may possess distinct continuum descriptions, depending on the scaling of $\{g_i,v_{i,i+1}\}$, and possibly different classes of mobile excitations, as discussed in a related context in \cite{Razamat:2021jkx}. 

Third, the present iterative rule generates a linear quiver and hence a single emergent dimension. More generally, however, the structure uncovered here suggests that iterative gauging is naturally organized by an underlying graph structure. The fields $A_i$ and $H_{ij}$ are naturally associated to the vertices and edges of a graph, respectively, and the resulting action coincides with the dual form of the deconstruction theory defined on that graph. The simplest one-dimensional construction discussed in this note corresponds to the special case of a chain, while more general graphs may describe discretizations of higher-dimensional geometries. It would be interesting to investigate whether generalized iterative gauging procedures can naturally generate such graph structures and to understand the resulting notions of emergent geometry. 

Fourth, one may consider a topological limit $g,v\rightarrow \infty$ in which the kinetic terms are removed, leaving only the BF couplings in \eqref{SAH}. It would be interesting to investigate the extent to which the resulting theory is sensitive to the topology of the underlying graph and, in the continuum limit, to that of the deconstructed geometry.

The construction in Ref.\cite{Rubio:2024rau} is formulated in an Abelian setting. Likewise, our procedure relies on the Abelian nature of the gauge groups in the quiver theory. As such, it can be regarded as identifying the Abelian skeleton of the deconstructed theory. This is somehow reminiscent of \cite{Franco:2022ziy}, where the sector containing the 1-form symmetry along the deconstructed dimensions sits in the Abelian part of the gauge group. It would be very interesting to understand whether a corresponding iterative-gauging description exists for genuinely non-Abelian degrees of freedom and, more generally, how the full non-Abelian dynamics of deconstruction is encoded in the present framework. Recent developments suggest that non-Abelian generalizations of iterative gauging may be closely connected to categorical or non-invertible symmetries~\cite{Blanik:2025nonabelian}.

More broadly, our result suggests that deconstruction is a unifying concept that connects the higher-form symmetry language of modern quantum field theory to the extra-dimensional intuition of the 2000s. It would be interesting to explore whether other instances of emergent dimensions in the recent higher-form symmetry literature admit analogous deconstruction interpretations.

%===================================================
\section*{Acknowledgments}
%===================================================

The authors thank the Simons Center for Geometry and Physics for hospitality during the \textit{2026 Simons Physics Summer Workshop}. D.R.G. would like to thank the organizers and the participants of the \textit{Developments on Generalized Symmetries of Quantum Systems} at the University of Murcia for very stimulating discussions leading to this project. S.F. is supported by the U.S. National Science Foundation grant PHY-2412479. D.R.G is supported in part by the Spanish national grant MCIU-22-PID2021-123021NB-I00.

\end{document}